\documentclass[12pt]{article}

\usepackage{apalike}

\usepackage[dvips]{graphicx}
\usepackage{color}
\usepackage{filecontents}

\usepackage{breakcites}

\newcommand{\mc}{\mathcal}
\newcommand{\real}{\mathbb{R}}

\newcommand{\map}[3]{#1: #2 \rightarrow #3}

\usepackage{amssymb}
\usepackage{amsfonts}
\usepackage{amsmath}
\usepackage[linesnumbered,ruled,vlined]{algorithm2e}
\usepackage{color}

\usepackage{caption}
\usepackage{subcaption}

\usepackage{color}

\usepackage{times}
\usepackage{graphicx}
\usepackage{textgreek}
\usepackage{gensymb}
\usepackage{setspace}

\usepackage{color}

\usepackage{booktabs}

\newcounter{lastnote}

\title{Network Controllability in the Inferior Frontal Gyrus Relates to Controlled Language Variability and Susceptibility to TMS}

\author{John D. Medaglia,$^{1,2\ast}$ Denise Y. Harvey,$^{2,3}$ Nicole White,$^{2}$,\\ Apoorva Kelkar,$^{1}$ Jared Zimmermann,$^{4}$ Danielle S. Bassett,$^{2,5,6}$\\ Roy H. Hamilton$^{2}$\\
	\normalsize{$^{1}$ Department of Psychology, Drexel University}\\
	\normalsize{Philadelphia, PA, 19104, USA}\\
	\normalsize{$^{2}$ Department of Neurology, University of Pennsylvania}\\
	\normalsize{Philadelphia, PA, 19104, USA}\\
	\normalsize{$^{3}$ Moss Rehabilitation Research Institute}\\
	\normalsize{Elkins Park, PA, 19027, USA}\\
	\normalsize{$^{4}$ Department of Neuroscience, University of Pennsylvania}\\
	\normalsize{Philadelphia, PA, 19104, USA}\\
	\normalsize{$^{5}$ Department of Electrical and Systems Engineering, University of Pennsylvania}\\
	\normalsize{Philadelphia, PA, 19104, USA}\\
	\normalsize{$^{6}$ Department of Bioengineering, University of Pennsylvania}\\
	\normalsize{Philadelphia, PA, 19104, USA}\\
	\\
	\normalsize{$^\ast$ To whom correspondence should be addressed; E-mail:  johnmedaglia@gmail.com.}
}

\makeatletter
\let\@internalcite\cite
\def\cite{\def\citeauthoryear##1##2{##1, ##2}\@internalcite}

\makeatother

\begin{document} 
	\maketitle
	\noindent 	Number of pages: 38\\
	Number of figures: 7\\
	Number of tables: 4\\
	Abstract words: 217\\
	Introduction words: 632\\
	Discussion words: 1391\\
	
	\doublespacing
	
	\maketitle
	\newpage
	\begin{abstract}
		In language production, humans are confronted with considerable word selection demands. Often, we must select a word from among similar, acceptable, and competing alternative words in order to construct a sentence that conveys an intended meaning. In recent years, the left inferior frontal gyrus (LIFG) has been identified as critical to this ability. Despite a recent emphasis on network approaches to understanding language, how the LIFG interacts with the brain's complex networks to facilitate controlled language performance remains unknown. Here, we take a novel approach to understand word selection as a network control process in the brain. Using an anatomical brain network derived from high-resolution diffusion spectrum imaging (DSI), we computed network controllability underlying the site of transcranial magnetic stimulation in the LIFG between administrations of two word selection tasks. We find that a statistic that quantifies the LIFG's theoretically predicted control of difficult-to-reach states explains vulnerability to TMS in language tasks that vary in response (cognitive control) demands: open-response (word generation) vs. closed-response (number naming) tasks. Moreover, we find that a statistic that quantifies the LIFG's theoretically predicted control of communication across modules in the human connectome explains TMS-induced changes in open-response language task performance only. These findings establish a link between network controllability, cognitive function, and TMS effects.
	\end{abstract}
	\newpage
	
	\section*{Significance Statement}
	This work illustrates that network control statistics applied to anatomical connectivity data demonstrate relationships with cognitive variability during controlled language tasks and TMS effects. 
	\newpage
	
	\section*{Introduction}
	Effective verbal communication depends on the ability to retrieve and select the appropriate words that correspond to a speaker's intended meaning. Often, the opportunity to select among several appropriate words challenges the speaker. Prior evidence in cognitive neuroscience indicates that the left inferior frontal gyrus (LIFG) supports verbal selection \cite{botvinick2001conflict,moss2005selecting,nelson2009mapping,snyder2011choosing,thompson1997role,thompson1999effects,thompson2006resolving,tippett2004selection}, and potentially a more domain-general role in selection in the context of competing representations \cite{fedorenko2014reworking}. Notably, the position of the LIFG in the brain's distributed anatomical networks is not unique to classically described language systems. Rather, it is positioned to mediate between several systems in the frontal associative, motor, insular, and temporal cortices as well as the basal ganglia \cite{saur2008ventral}. This evidence suggests that the participation of the LIFG in language function must operate in the context of many processing demands in the brain's distributed circuits.
	
	While controlled language function is thought to be a network-level process \cite{doron2012dynamic,chai2016functional,fedorenko2014reworking}, putative mechanisms of this process in the context of the brain's complex structural architecture remain unclear. Recent theoretical work in network control theory, an emerging area in engineering, provides one such mechanism. Network control theory (NCT) is the study of how to design control strategies for networked systems \cite{ruths2014control}, in which a set of nodes are connected by edges, whereby a particular dynamic process occurs atop those edges. In the context of the brain, this suggests that brain regions (nodes) are predisposed to drive or modulate neurophysiological dynamics in a manner consistent with their specific topological role in brain networks constructed from white matter tractography. Variability in nodes' ability to drive the network into different trajectories may account for performance variability in control-demanding tasks \cite{gu2015controllability}, and therefore the system's susceptibility to perturbation via transcranial magnetic stimulation (TMS). However, a mechanistic network control role for the LIFG in language has not been experimentally tested.
	
	Here we test whether NCT is a putative mechanism for language control by asking whether the theoretically predicted (i.e., mathematically derived) network control features of brain regions are related to cognitive performance word retrieval tasks with varying cognitive control (and thus LIFG) involvement by virtue of their response demands \cite{hoffman2010ventrolateral}. In particular, we focus on open-ended semantic tasks, where participants can choose one of several appropriate words to complete the task \cite{botvinick2001conflict}. This contrasts with a closed-ended task -- number naming -- that requires word retrieval but has only one correct response. We posit that language performance relates to the ability of the LIFG to control activity across human anatomical brain networks. In particular we anticipate dissociable contributions of network control roles to open- \cite{jefferies2006semantic,noonan2013going,hoffman2013semantic,jefferies2013neural} and closed- language tasks, which have different response retrieval and selection demands mediated by the \emph{pars triangularis} \cite{hoffman2010ventrolateral}.
	
	To assess this view, we focus on two distinct network control features known as modal controllability (the theoretical ability of a node -- here, a brain region -- to drive a network into difficult-to-reach states) and boundary controllability (the theoretical ability of a node to steer the system into states where modules are either coupled or decoupled). Boundary controllability in the LIFG may represent inter-system coordination required for effective language production, such as retrieving and selecting a single word in the face of competing, alternative words. Modal controllability in the LIFG may represent the recruitment of specific task-related states necessary for producing an accurate response (i.e., retrieving and selecting specific words according to varying task and/or response demands). We hypothesize that local inhibition via brain stimulation \cite{benali2011theta,huang2005theta} will allow us to distinguish network control roles corresponding to different cognitive roles of the \emph{pars triangularis} \cite{hoffman2010ventrolateral} during open- and closed- language tasks.
	
	\section*{Materials \& Methods}
	\subsection*{Overview of Methods}
\begin{figure}[h!]
		\centerline{\includegraphics[width=6.5in]{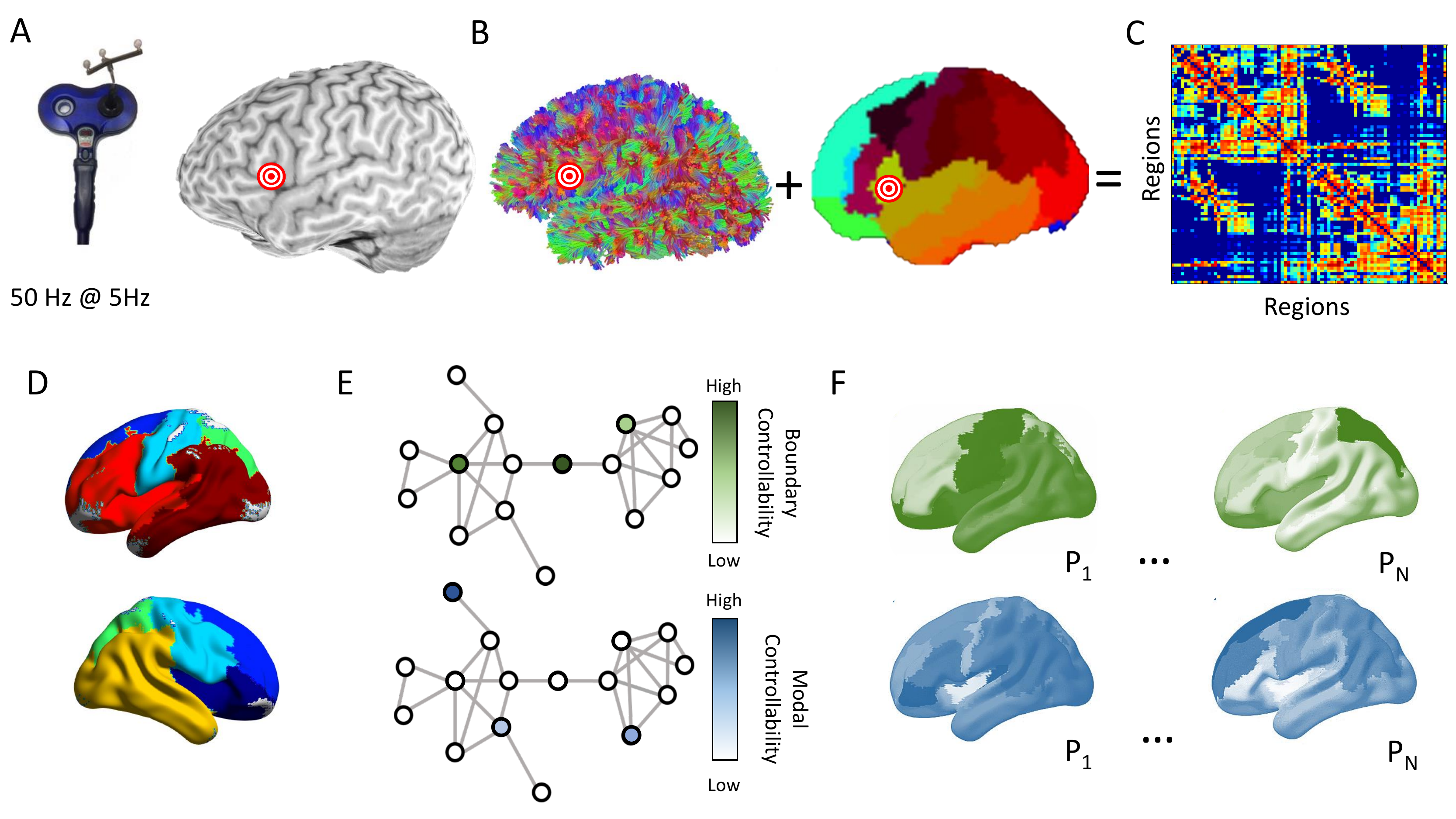}}
		\caption{\textbf{Overview of Methods} \emph{(A)} Continuous theta burst stimulation was administered to each subject's \emph{pars triangularis}. \emph{(B)} Diffusion tractography was computed for each subject. A cortical parcellation was registered to each individual's structural T1 image to identify anatomical divisions. \emph{(C)} A region x region structural adjacency matrix was constructed representing the streamline counts between pairs of regions. \emph{(D)} We applied a community detection algorithm to identify an initial consensus partiation on the average network across subjects. \emph{(E)} Modal and boundary controllability were computed for each node (brain region) in the network for each individual.  Each node received a rank representing its strength of control in the network heirarchy within the individual. \emph{(F)} Maps representing the variability in modal controllability (top) and boundary controllability (bottom). $P_{1...N}$ represent different participants. The relationship between controllability values at the LIFG stimulation site and task response times before and after stimulation were examined using mixed effects models.}\label{methods}
	\end{figure}

	To address our hypotheses, we administered a form of noninvasive brain stimulation (transcranial magnetic stimulation; TMS) to a region within the left inferior frontal gyrus (\emph{pars triangularis}) in each of 19 healthy adult subjects between repeated administrations of two language tasks with open-ended selection demands and one number naming task with a single appropriate response for comparison. We also administered sham TMS to the vertex in each of 9 healthy adult subjects between repeated administrations of the same tasks. Then, we constructed structural brain networks from diffusion spectrum imaging (DSI) data acquired for each subject (Methods, Fig.~\ref{methods}A). Each network contained 111 brain regions defined by the Lausanne anatomical parcellation and cerebellum (Fig.~\ref{methods}B), and each pair of regions was connected by an edge weighted by the number of streamlines linking those regions (Fig.~\ref{methods}C). We defined a simplified model of brain dynamics and simulated network control to quantify modal and boundary controllability (Fig.~\ref{methods}D). 
	
	\subsection*{Subjects}
	Twenty-eight healthy individuals (mean age = 25.4, St.D. = 4.5, 16 female) were scanned on a 3T Prisma scanner at the University of Pennsylvania in the present study. All procedures were approved in a convened review by the University of Pennsylvania's Institutional Review Board and were carried out in accordance with the guidelines of the Institutional Review Board/Human Subjects Committee, University of Pennsylvania. All participants volunteered with informed consent in writing prior to data collection.
	
	\section*{Neuroimaging: Diffusion Tractography}
	Diffusion spectrum images (DSI) were acquired for a total of 28 subjects along with a T1-weighted anatomical scan at each scanning session. We followed a parallel strategy for data acquisition and construction of streamline adjacency matrices as in previous work applying network controllability statistics in human diffusion imaging networks \cite{gu2015controllability}. DSI scans sampled 257 directions using a Q5 half-shell acquisition scheme with a maximum $b$-value of 5,000 and an isotropic voxel size of 2.4 mm. We utilized an axial acquisition with the following parameters: repetition time (TR) = 5 s, echo time (TE) = 138 ms, 52 slices, field of view (FoV) (231, 231, 125 mm). 
	
	DSI data were reconstructed in DTI Studio (www.dsi-studio.labsolver.org) using $q$-space diffeomorphic reconstruction (QSDR) \cite{yeh2011estimation}. QSDR first reconstructs diffusion-weighted images in native space and computes the quantitative anisotropy (QA) in each voxel. These QA values are used to warp the brain to a template QA volume in Montreal Neurological Institute (MNI) space using the statistical parametric mapping (SPM) nonlinear registration algorithm. Once in MNI space, spin density functions were again reconstructed with a mean diffusion distance of 1.25 mm using three fiber orientations per voxel. Fiber tracking was performed in DSI Studio with an angular cutoff of 35$^\circ$, step size of 1.0 mm, minimum length of 10 mm, spin density function smoothing of 0.0, maximum length of 400 mm and a QA threshold determined by DWI signal in the colony-stimulating factor. Deterministic fiber tracking using a modified FACT algorithm was performed until 1,000,000 streamlines were reconstructed for each individual. Streamlines were initiated at the voxel level to utilize the initial resolution of the diffusion images.
	
	Anatomical (T1) scans were segmented using FreeSurfer \cite{fischl2012freesurfer} and parcellated using the connectome mapping toolkit \cite{cammoun2012mapping}. A parcellation scheme including $n=111$ regions was registered to the B0 volume from each subject's DSI data. The B0 to MNI voxel mapping produced via QSDR was used to map region labels from native space to MNI coordinates. To extend region labels through the grey-white matter interface, the atlas was dilated by 4 mm \cite{cieslak2014local}. Dilation was accomplished by filling non-labelled voxels with the statistical mode of their neighbors' labels. In the event of a tie, one of the modes was arbitrarily selected. Each streamline was labelled according to its terminal region pair. From these data, we constructed a structural connectivity matrix, $\mathbf{A}$ whose element $A_{ij}$ represented the number of streamlines connecting different regions, divided by the sum of volumes for regions $i$ and $j$ \cite{Hagmann2008}. Notably, there are numerous free parameters in diffusion tractography, image parcellation, and graph representations of anatomical connectivity (e.g., weighted \emph{versus} binarized --or unweighted-- graphs). Here, we aimed to remain consistent with other work that examined controllability in weighted streamline networks \cite{gu2015controllability,betzel2016optimally,tang2017developmental,Bassett2006}. Controllability profiles are similar when using probabalistic tractography or weight networks by fractional anisotropy \cite{tang2017developmental}.
	
	\subsection*{Cognitive Testing}
\begin{figure}[h!]
		\centerline{\includegraphics[width=6.5in]{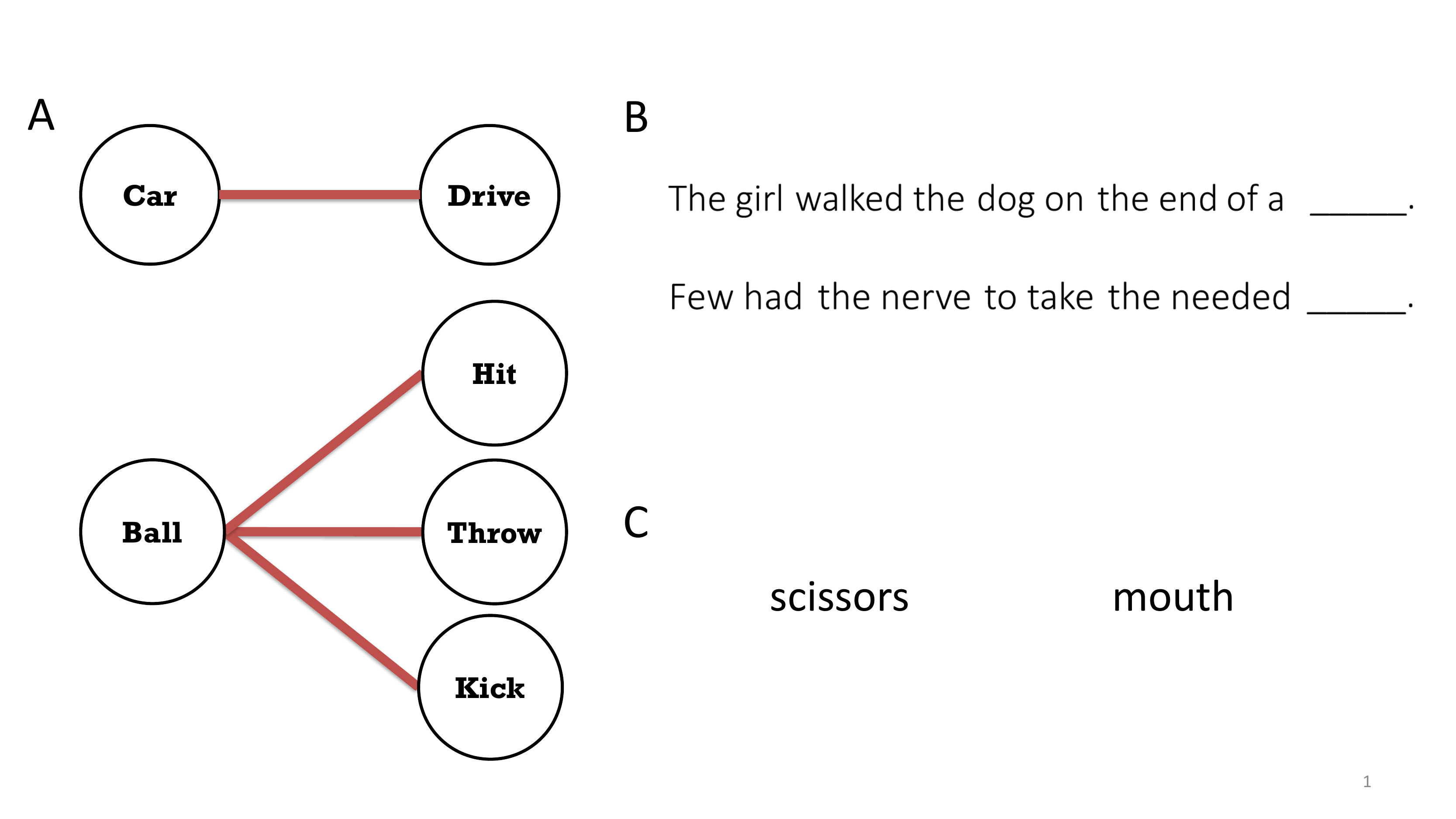}}
		\caption{\textbf{Overview of Tasks} \emph{(A)} High verbal selection demands are introduced when a cuing noun is associated with multiple appropriate words (here, verbs). \textit{Top}: Example of a stimulus-response pairing with low selection demands. \textit{Bottom}: a stimulus-response pairing with high selection demands. \emph{(B)} Example items from the sentence completion task. Participants were asked to provide an appropriate noun at the end of the sentence. \textit{Top}: This item has a low selection demand because ``leash" is easily and dominantly recalled in the context of this sentence. \textit{Bottom}: This item has a high selection demand because several alternate words may be appropriate to complete the sentence. \emph{(C)} Example items from the verb generation task. Participants were asked to provide an appropriate verb associated with the noun. \textit{Left}: This item has a low selection demand because ``cut" is the most dominant verb associated with ``scissors". \textit{Right}: This item has a high selection demand because several verbs are highly associated with the mouth, such as ``eat", ``talk", and ``kiss".}\label{tasks}
	\end{figure}

	Participants performed two open-ended language tasks and one closed-ended number naming task (See Fig.~\ref{tasks}). 
	
	The language tasks included a verb generation task and a sentence completion task (e.g., \cite{snyder2008so}). For the verb generation task, subjects were instructed to generate the first verb that came to mind when presented with a noun stimulus (e.g., ``cat''). The verb could be either something the noun does (e.g., ``meow'') or something you do with it (e.g., ``feed''). Response times (RTs) were collected from the onset of the noun cue to the onset of the verb response. For the sentence completion task, participants were presented with a sentence, such as ``They left the dirty dishes in the -----.'', and were instructed to generate a single word that appropriately completes the sentence, such as ``sink''. Words in the sentences were presented serially in 1s segments consisting of one or two words. RTs were computed as the latency between the onset of the last segment, which always contained a two-word segment (i.e., a word and an underline), and the onset of the participant's response. For all items in the sentence completion task, items in the high vs. low selection demand conditions were matched on retrieval demands (association strength) \cite{snyder2008so}. For both language tasks, each trial began with the presentation of a fixation point (+) for 500 ms, followed by the presentation of the target stimulus, which remained on the screen for 10 s until the subject made a response. Subjects were given an example and five practice trials in the first administration of each language task (i.e., before TMS), and were reminded of the instructions before performing the task a second time (i.e., after TMS). In each of the before and after TMS conditions, subjects completed 50 trials for a total of 100 trials. 
	
	The items for the verb generation task were identical to those used in \cite{snyder2011choosing} and the items for the sentence completion task were those from \cite{snyder2014opposite}. The difficulty of items was sampled to cover a distribution of values computed via latent semantic analysis (LSA) applied to corpus data. In particular, items were sampled to represent a range of LSA entropy and LSA association strength \cite{snyder2008so}, which represent the selection and retrieval demands of each item, respectively \cite{snyder2008so}. An LSA association value of 0 means that the cue word or sentence is not strongly associated with any word in particular, whereas a value of 1 means that the cue word or sentence is strongly associated with at least one word, implying that it is easy to retrieve. An LSA entropy value of 0 indicates that the word is not related to any words, whereas higher values indicate higher relatedness to many words, which theoretically increases competition among appropriate words \cite{snyder2008so}. 
	
	The comparison task requiring closed-ended responses was a number naming task where participants produced the English names for strings of Arabic numerals presented on the screen. On each trial, a randomized number (from tens of thousands to millions; e.g., 56395, 614592, 7246856) was presented in black text on a white background. The numbers were uniformly distributed over three lengths (17 per length for each task administration). The position of items on the screen was randomized between the center, left, and right of the screen to reduce the availability of visual cues to number length and syntax \cite{snyder2008so}. RTs were collected from the onset of the stimulus presentations to the onset of the subject's response. The number appeared in gray following the detection of a response (i.e., voice key trigger), and remained on the screen thereafter to reduce the working memory demands required for remembering the digit string. At the start of the experiment, subjects performed 50 trials of the number naming task to account for initial learning effects \cite{snyder2008so}. Prior to performing the task for the first time, subjects were given an example and five practice trials, and were later reminded of the instructions before performing the task a second (i.e., before TMS) and a third time (i.e., after TMS). In each of the before and after TMS conditions, subjects completed 51 trials for a total of 102 experimental trials.
	
	Verbal responses for all tasks were collected from a computer headset microphone. The microphone was calibrated to reduce sensitivity to environment background noise prior to the collection of data for each session such that the recording software was not triggered without clear verbalizations. List order (before or after TMS) was counterbalanced across participants. Item presentation order within each task was fully randomized across participants.
	
	\subsection*{Transcranial Magnetic Stimulation}
	The Brainsight system (Rogue Research, Montreal) was used to co-register MRI data with the location of the subject and the TMS coil. The stimulation site was defined as the posterior extent of the \emph{pars triangularis} in each individual subject's registered T1 image. A Magstim Super Rapid$^{2}$ Plus$^{1}$ stimulator (Magstim; Whitland, UK) was used to deliver cTBS via a 70 mm diameter figure-eight coil. To calibrate the intensity of stimulation, cTBS was delivered at 80\% of each participant'™s active motor threshold \cite{huang2005theta}. Each participant's threshold was determined prior to the start of the experimental session using a standard up-down staircase procedure with stimulation to the motor cortex (M1).
	
	\section*{Mathematical Models}
	\subsection*{Network Control Theory}
	We follow a previous application of network control theory in diffusion weighted imaging data as the basis for our examination of controllability and cognitive control. We briefly describe the mathematical basis for the approach taken here. For a full discussion of structural network controllability in the context of diffusion weighted imaging networks, see \cite{gu2015controllability}. For a full discussion of the mathematical basis for structural network controllability see \cite{liu2011controllability,ruths2014control,pasqualetti2014controllability}.
	
	Our ability to understand neural systems is fundamentally related to our ability to control them \cite{Schiff2012}. Network control theory is a branch of traditional control theory in engineering that examines how to control a system based on the pattern of links between its components, and based on a model of the system's dynamics. Here, we interpret the word \emph{control} to mean perturbing communication in an anatomical brain network. To apply a network control perspective, we require (i) knowledge of the network connectivity linking system components, and (ii) knowledge regarding how system components function, i.e., their \emph{dynamics}, rather than simply a descriptive statistics of the network's architecture. In contrast to traditional graph theory, network control theory offers mechanistic predictors of network dynamics. The use of mechanistic models allows us enrich descriptive approaches to examine the human connectome \cite{medaglia2015cognitive} with statistics that explicitly include a dynamic model.
	
	Mathematically, we can study the controllability of a networked system by defining a network represented by the graph $\mc G = (\mc V, \mc E)$, where $\mc V$ and $\mc E$ are the vertex and edge sets, respectively. Let $a_{ij}$ be the weight associated with the edge $(i,j) \in \mc E$, and define the \emph{weighted adjacency matrix} of $\mc G$ as $A = [a_{ij}]$, where $a_{ij} = 0$ whenever $(i,j) \not\in \mc E$. We associate a real value (\emph{state}) with each node, collect the node states into a vector (\emph{network state}), and define the map $\map{x}{\mathbb{N}_{\ge 0}}{\mathbb{R}^n}$ to describe the evolution (\emph{network dynamics}) of the network state over time. Given the network and node dynamics, we can use network control theory to quantitatively examine how the network structure relates to the types of control that nodes can exert.
	
	\subsection*{Dynamic Model of Neural Processes} We begin with an analogous approach to prior work \cite{gu2015controllability}. We define structural brain networks by subdividing the entire brain into anatomically distinct brain areas (network nodes) in a commonly used anatomical atlas \cite{Hagmann2008}. Consistent with prior work \cite{Bassett2011,Hermundstad2013,Hermundstad2014,gu2015controllability}, we connect nodes by the number of white matter streamlines identified by a commonly used deterministic tractography algorithm (for details on the tractography implementation, see \cite{Cieslak2014}). This procedure results in sparse, weighted, undirected structural brain networks for each subject. Properties of this network include high clustering, short path length, and strong modularity, consistent with prior studies of similar network data \cite{Bassett2011,Hagmann2008}. The definition of structural brain networks based on tractography data in humans follows from our primary hypothesis that control features of neural dynamics are in part determined by the structural organization of the brain's white matter tracts.
	
	To define the dynamics of neural processes, we draw on prior models linking structural brain networks to resting state functional dynamics \cite{honey2009predicting,honey2010can,abdelnour2014network}. Although neural activity evolves through neural circuits as a collection of \emph{nonlinear} dynamic processes, these prior studies have demonstrated that a significant amount of variance in neural dynamics as measured by resting state fMRI can be predicted from simplified \emph{linear} models. Based on this literature, we employ a simplified noise-free linear discrete-time and time-invariant network model: 
	
	\begin{equation}\label{eq: linear network}
	\mathbf{x} (t+1) = \mathbf{A} \mathbf{x}(t) + \mathbf{B}_{\mc K} \mathbf{u}_{\mc K} (t),
	\end{equation}
	
	\noindent	where $\map{\mathbf{x}}{\real_{\ge 0}}{\real^N}$ describes the state (e.g., a measure of the electrical charge, oxygen level, or firing rate) of brain regions over time, and $\mathbf{A} \in \real^{N \times N}$ is a symmetric and weighted adjacency matrix. In this case, we construct a weighted adjacency matrix whose elements indicate the number of white matter streamlines connecting two different brain regions -- denoted here as $i$ and $j$ -- and we stabilize this matrix by dividing by the mean edge weight. While the model employed above is a discrete-time system, we find that the controllability Gramian is statistically similar to that obtained in a continuous-time system \cite{gu2015controllability}.
	
	The diagonal elements of the matrix $\mathbf{A}$ satisfy $A_{ii}=0$. The input matrix $\mathbf{B}_{\mc K}$ identifies the control points $\mc K$ in the brain, where $\mc K = \{k_1, \dots, k_m \}$ and \begin{align}\label{eq: B}
	B_{\mc K} =
	\begin{bmatrix}
	e_{k_1} & \cdots & e_{k_m}
	\end{bmatrix},
	\end{align}
	and $e_i$ denotes the $i$-th canonical vector of dimension $N$. The input $\map{\mathbf{u}_{\mc K}}{\real_{\ge  0}}{\real^m}$ denotes the control energy.
	
	\subsection*{Network Controllability} To study the ability of a certain brain region to influence other regions in arbitrary ways we adopt the control theoretic notion of \emph{controllability}. Controllability of a dynamical system refers to the possibility of driving the state of a dynamical system to a specific target state by means of an external control input \cite{REK-YCH-SKN:63}. In the current paper, we follow the procedures applied in \cite{gu2015controllability} and focus on two network controllability statistics: \emph{modal} and \emph{boundary} controllability. It is critical to note that the controllability statistics examined here are based on linear discrete time dynamics, but they have also recently been extended to nonlinear models of dynamics. In simulation studies, linear controllabilty statistics were related to predicted effects in dynamics simulated using Wilson-Cowan \cite{muldoon2016stimulation} oscillators in anatomical brain networks. In addition, they predicted topological changes in network dynamics simulated using Kuramoto oscillators \cite{tiberi2017synchronization}. Thus, we focus on mathematically well defined linear control statistics due to their parsimony and pragmatic utility in applied contexts, such as neuromodulation research.
	
	\paragraph{Boundary Controllability.}
	\emph{Boundary controllability}, a metric developed in network control theory, quantifies the role of a network node in controlling dynamics between modules in hierarchical modular networks \cite{pasqualetti2014controllability}. Boundary controllability identifies brain areas that can steer the system into states where different cognitive systems are either coupled or decoupled. A region's boundary controllability describes its theoretical ability to regulate the extent to which it can drive major networks to increase or decrease communication with one another. High boundary controllers are conceptually akin to the ``gatekeepers" of communication between major brain networks. Here, we apply a similar approach to that taken in \cite{gu2015controllability} to quantify boundary controllability in our diffusion tractography networks and associate controllability variability with cognitive performance. Specifically, we partition the brain into \emph{modules} by maximizing the modularity quality function \cite{Newman2006} using a Louvain-like \cite{Blondel2008} locally greedy algorithm \cite{Jutla2011}. Because the modularity quality function has many near-degeneracies, we perform the optimization algorithm multiple times \cite{Good2010}. We observed that the mean partition similarity was high and the variance of the partition similarity was low for a value of $\gamma$ at $1.6$ (mean z-Rand score = 60.4, standard deviation = 3.7), which is within the range of stable partitions found in prior analyses of diffusion spectrum imaging data \cite{gu2015controllability}. We therefore used the consensus partition at $\gamma = 1.6$ for the remainder of the analysis in this study. High ranking boundary controllers are identified as the highest ranking set of boundary regions, and remaining boundary regions are found within modules in the network. For our regression analyses, the high ranking set introduces a ceiling effect and is conceptually distinct from all other controllers; therefore, they were binarized for our multilevel regression analyses.
	
	\paragraph{Modal Controllability.}
	\emph{Modal} controllability refers to the ability of a node to control each evolutionary mode of a dynamical network \cite{AMAH-AHN:89}, and can be used to identify the \emph{least controllable} state from a set of control nodes. Modal controllability is computed from the eigenvector matrix $V = [v_{ij}]$ of the network adjacency matrix $\mathbf{A}$. By extension from the PBH test \cite{TK:80}, if the entry $v_{ij}$ is small, then the $j$-th mode is poorly controllable from node $i$. Following \cite{FP-SZ-FB:13q}, we define $\phi_i =\sum_{j =1}^N (1 - \lambda_j^2 (A)) v_{ij}^2$ as a scaled measure of the controllability of all $N$ modes $\lambda_1 (A),\dots, \lambda_N(A)$ from the brain region $i$. Regions with high modal controllability are able to control all the dynamic configurations of the network, and hence to drive the dynamics towards hard-to-reach configurations. A brain region's modal controllability describes its theoretical ability to drive the brain into states that are difficult to reach. Dynamically, these states typically involve the activation of a few, specific regions in the network. High modal controllers are conceptually akin to dynamic ``specialists" driving specific, otherwise unachievable states.
	
	\paragraph{Node centrality and controllability.}
	There is a key distinction between the controllability statistics and more commonly applied node centrality statistics. While more common node centrality measures are used to make inferences about the role of brain regions in information processing \cite{Rubinov2010}, network controllability statistics explicitly encode dynamics. This means that they support a theoretical inference about the role of brain regions in controlling states. One crucial value of applying network control theory in neural stimulation is that it paves the way for explicit connections to systems engineering and control theory, extending past passive analysis of network topology.
	
	In the current data, modal controllability in the LIFG is negatively but imperfectly correlated with node strength (the sum of edge weights emanating from LIFG; $R = -0.62, p = 2.34 \times 10^{-4}$), not correlated with node betweenness centrality (the number of shortest paths through the network that pass through the LIFG; $R = -0.03, p = 0.87$) and not correlated with node closeness centrality (the sum of the length of shortest paths between the LIFG and all other regions; $R = 0.03, p = 0.85$). Boundary controllability is not significantly correlated with node strength ($R = 0.23, p = 0.23$) or node betweenness centrality ($R = 0.14, p = 0.45$) and is moderately positively correlated with node closeness centrality ($R = 0.34, p = 0.06$).
	
	These findings indicate that in addition to supporting a theoretical inference in network control mediated by the LIFG, control roles are distinguishable from basic measures of centrality calculated from static network topology.
	
	\subsection*{Examining the Relationship Between Controllability, Cognition, and TMS effects}
	Analyses were conducted using multilevel modeling with maximum-likelihood estimation \cite{baayen2008mixed} implemented in the lme4 v.1.1-9  \cite{bates2014lme4} package of R version 3.2.1 \cite{r2016software}. This technique allows for a classical regression analysis to be performed on repeated measures data by accounting for the non-independence of observations collected from each participant in a within-subjects design, without resorting to computing separate regression equations for each subject \cite{lorch1990regression,baayen2008mixed,baayen2008analyzing}. Multilevel modeling also accounts for violations of the sphericity assumption by modeling heteroskedasticity in the data when necessary, improving statistical power over other methods commonly employed for analyzing repeated-measures data. We excluded from analyses trials on which participants responded incorrectly (i.e., semantic and paraphasic errors, hesitations, or false starts) and experimenter error/equipment failures (such as false triggers for voice recording), constituting a mean of 6.66\% and 3.68\% of all trials, respectively. Response times (RTs) in open- and closed-ended tasks were log-transformed due to non-normal distribution of raw RTs.
	
	To independently isolate relationships between open and closed task performance as a function of boundary or modal controllability and their potential modulation via TMS, we computed 4 separate mixed models (i.e., 2 each assessing boundary and modal controllability on open- and closed-task performance). For models assessing boundary controllability, fixed effects included the within-subjects factor of Session (before vs. after stimulation) and the between-subjects factors of TMS (active vs. sham) and LIFG boundary control (High vs. Low). Here, boundary controllability was binarized because nodes can be conceptually distinguished into those at the top of the modular heirarchy -- the highest level of intermodular interactions-- \emph{versus} all lower-ordered nodes. Thus, binarizing represents this distinction and removes the ceiling effect introduced by the high-rank boundary control values in the data. Separate models were then run with log-transformed RTs for the open-ended (i.e., verb generation and sentence completion) and closed-ended (i.e., number naming) tasks as the dependent variable.
	
	Models assessing modal control included the within-subjects factor of Session (before vs. after stimulation) and the between-subjects factor of TMS (active vs. sham); however, modal control was treated as a continuous variable due to its conceptually continuous nature and the continuous distribution of modal ranks in the current data. As in boundary control, we then ran separate models using log-transformed RTs for the open- and closed-ended tasks as the dependent variable. In all analyses, we attempted to fit models using maximal random effect structures, as this method increases model generalizability \cite{barr2013random}. Random effect structures were simplified as required for model convergence to reduce the likelihood of model over-fitting. Final models reported below included by-participant random intercepts and by-participant random slopes for the effects of Trial within Task per Session, which controlled for overall differences in mean RTs and fatigue across the course of each task within each experimental session. None of the final models reported produced convergence warnings indicative of model over-fitting.
	
	\subsection*{Code Availability}
	Code for calculating the controllability statistics examined here is available on Github (https://github.com/nangongwubu/Network-Controllability-Diagnostics).
	
	\section*{Results}
\begin{figure} 
		\centerline{\includegraphics[width=5in]{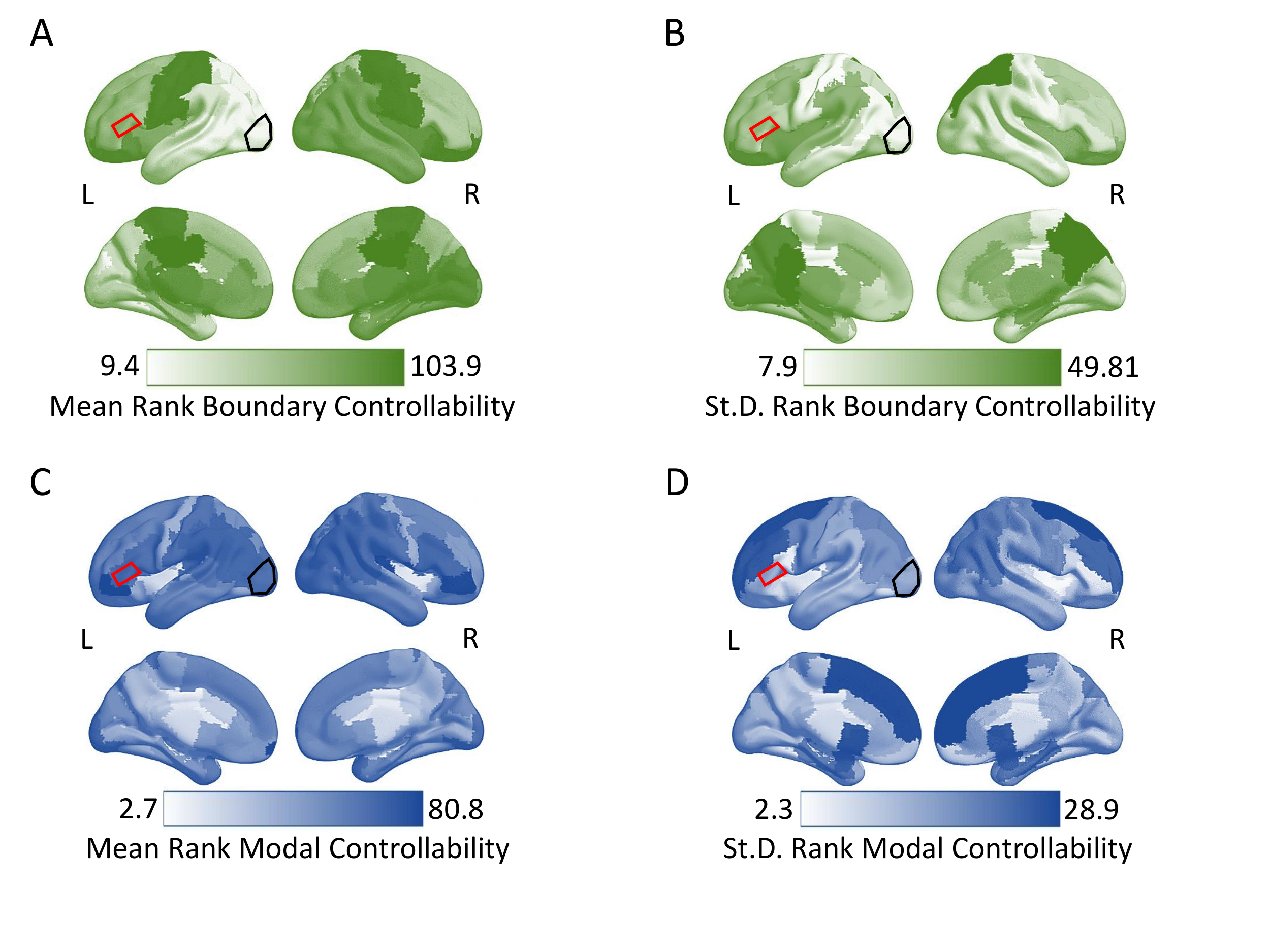}}
		\caption{\textbf{Mean and standard deviations (St.D.) of ranked boundary and modal controllability values.} Each colormap represents either the mean or St.D. of ranked controllability in brain regions across all subjects in the study. Darker colors represent either higher mean ranks or St.D. for that region.\emph{(A)} Mean ranked modal controllability across brain regions; \emph{(B)} standard deviation of ranked modal controllability values across brain regions; \emph{(C)} mean ranked boundary controllability across brain regions; \emph{(D)} standard deviation of ranked boundary controllability values across brain regions. The red outlined region is the LIFG, which is also the site of stimulation. The black outlined region is our lateral occipital site serving as a ``bottom up" control region in our analyses. At the \emph{pars triangularis} (our LIFG region), boundary controllability had a middle-range value with moderate variability relative to other regions in the brain. Modal controllability in the \emph{pars triangularis} was high with low variability relative to other regions. The top of each panel represents a lateral view of the brain, and the bottom represents a medial view. L = left, R = right.}\label{fig:controllabilityplot}
	\end{figure}

	For context, we computed the mean and standard deviation of the ranked boundary and modal controllability values for all regions across the brain. As illustrated in Fig.~\ref{fig:controllabilityplot}, LIFG boundary controllability values in our sample tended to lie in the middle range with moderate variability when compared to other brain regions, whereas modal controllability in the LIFG tended to have higher values with low variability when compared to other regions. To be sure that boundary and modal controllability capture unique topological control roles in the context of the brain's anatomical network, we assessed the relationship between these two metrics across subjects in the current sample. Here, modal and boundary controllability in the LIFG are weakly correlated ($R = -0.23, p = 0.22$), sharing only approximately 4\% variance. This observation suggests that boundary and modal controllability theoretically contribute to distinct network control processes. We therefore present the results in separate models to isolate the effects associated with each network control role in the LIFG.

	\clearpage
	\newpage
	
	We computed four multilevel regression models for the relationships between modal and boundary controllability at the LIFG and performance on the open- and closed-demands language tasks before and after active vs. sham stimulation. In our interpretation of the results, we restrict our focus to cTBS-related findings indicating that inhibitory stimulation of the LIFG modulates performance (i.e., before vs. after differences), when compared to sham stimulation, and that the degree to which the LIFG specializes in boundary or modal controllability predicts responsiveness to TMS. The rationale for this focus is twofold. First, performance changes before vs. after stimulation will reveal whether receiving active stimulation alters potential practice-related effects as established with sham stimulation. Second, demonstrating that active stimulation to the LIFG (compared to sham stimulation) affects performance based on the LIFG's control role in the network provides a strong test of this region's involvement in language tasks with open- and closed-demands, which in turn sheds light on the potential mechanisms of LIFG-mediated control.
	
\begin{figure} 
		\centerline{\includegraphics[width=4in]{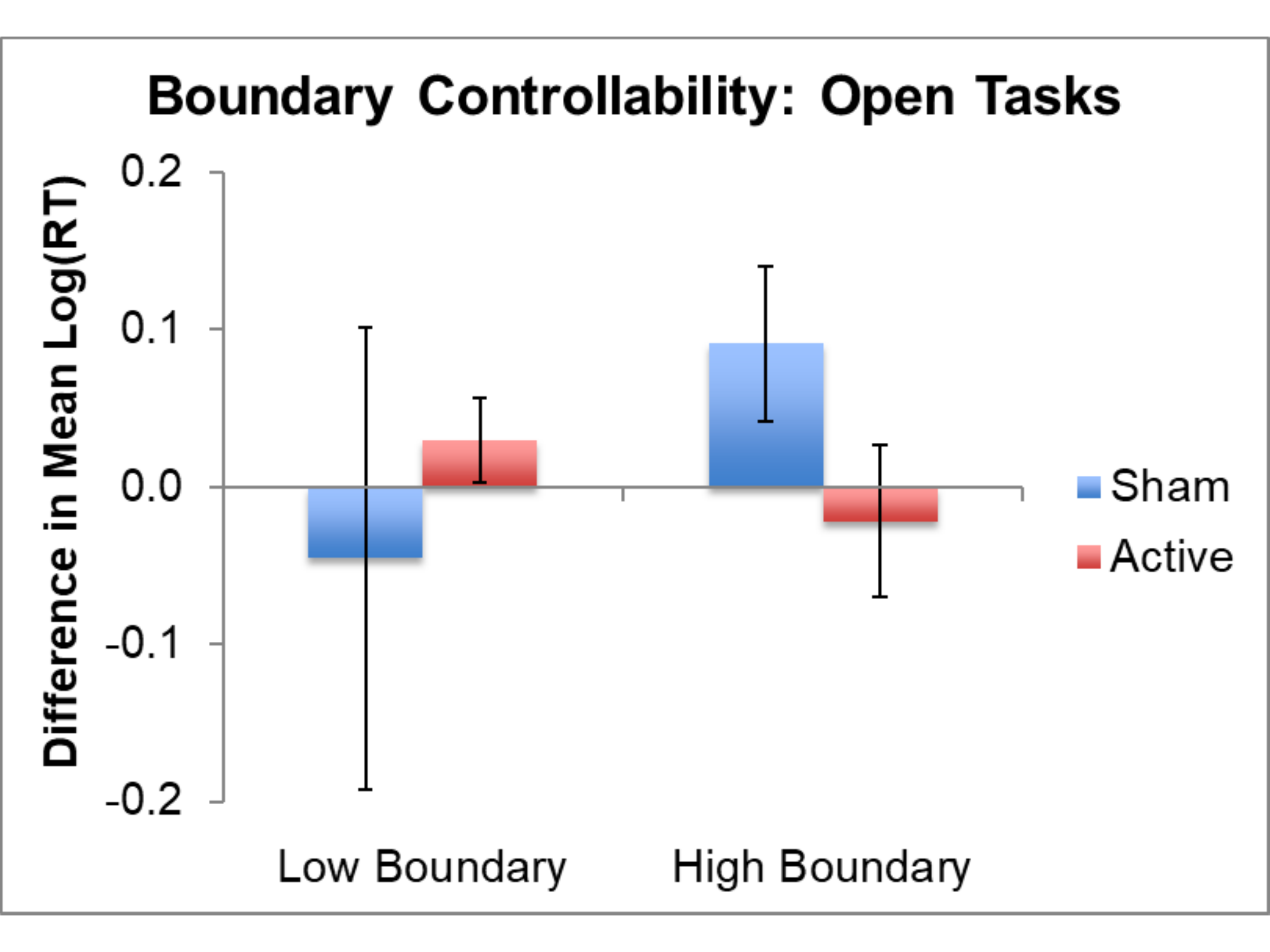}}
		\caption{\textbf{TMS effects in the significant interaction between TMS and boundary controllability in open language tasks.} Error bars represent standard error (SE) of the differences in mean log-transformed RTs. Lower values on the y-axis represent faster responses in the second session relative to the first session. See Table~\ref{table:boundopen} for full model results. Note that for transparency, the mean and SE of the log-transformed RTs do not represent model-adjusted values.}\label{fig:boundopen}
	\end{figure}

	\paragraph{\emph{Open-ended language task performance, TMS, and boundary and modal controllability.}} The model assessing the effect of boundary control on open-ended tasks revealed that cTBS of the LIFG modulated performance for individuals with high compared to low boundary controllability. Specifically, in the sham group, high boundary controllability was associated with longer RTs (henceforth reported as log-transformed values) after (when compared to before) sham stimulation (i.e., faster RTs before vs. after sham; mean (and SE) = 7.08 (.04) vs. 7.17 (.07), respectively), indexing a practice-related interference effect (i.e., slower to perform open-ended tasks a second time in the absence of real stimulation). Following cTBS, the practice-related interference effect associated with high boundary controllability was eliminated (i.e., similar RTs before and after cTBS; mean (and SE) = 7.16 (.06) and 7.14 (.07), respectively). For individuals with low boundary controllability, performance before vs. after stimulation did not differ as a function of active vs. sham stimulation. This observation suggests that the LIFG's theoretical role in integrating and segregating network communication may be important for retrieval and selection of one response when many appropriate alternatives compete. Inhibiting the LIFG may relieve practice-related interference effects when the LIFG is more heavily involved in a highly integrative role in the brain. See Table~\ref{table:boundopen} and Fig.~\ref{fig:boundopen}.
	
	\begin{table}[ht]
		
		\caption*{\textbf{Mixed effects model for the effects of boundary controllability, session, and TMS in the open tasks.}} 
		\begin{tabular}{l c c c c c c} \toprule
			
			{} & {Estimate} & {Std. Error} & {t-value} & {\emph{p}} \\ \midrule
			Intercept  & 7.167 & 0.049 & 145.975 & $<$0.001***\\
			Session    & -0.025& 0.020 & -1.232 & 0.212\\
			ActiveSham & -0.085& 0.074 & -1.147 & 0.252\\	  
			Boundary   & 0.023 & 0.068 & 0.189  & 0.850\\	 
			Session*Boundary   & 0.052 & 0.028 & 1.846  & 0.065\\	 
			Session*ActiveSham   & 0.112 & 0.031 & 3.658  & $<$0.001***\\
			Boundary*ActiveSham   & 0.355 & 0.136 & 2.603  & $<$0.009**\\	 
			Boundary*ActiveSham*Session   & -0.175 & 0.056 & -3.152 & $<$0.002**\\		 
			\bottomrule
			
		\end{tabular}
		\par
		\caption{ActiveSham = active transcranial magnetic stimulation effect relative to sham. Boundary = binarized boundary controllability effect. Session = effect of the second relative to the first session. *Denotes significance at $p < 0.05$, **denotes significance at $p < 0.01$, and ***denotes significance at $p < 0.001$.}
		\label{table:boundopen}
	\end{table}
	\newpage
	
	\clearpage
	\newpage
	
\begin{figure} 
		\centerline{\includegraphics[width=4in]{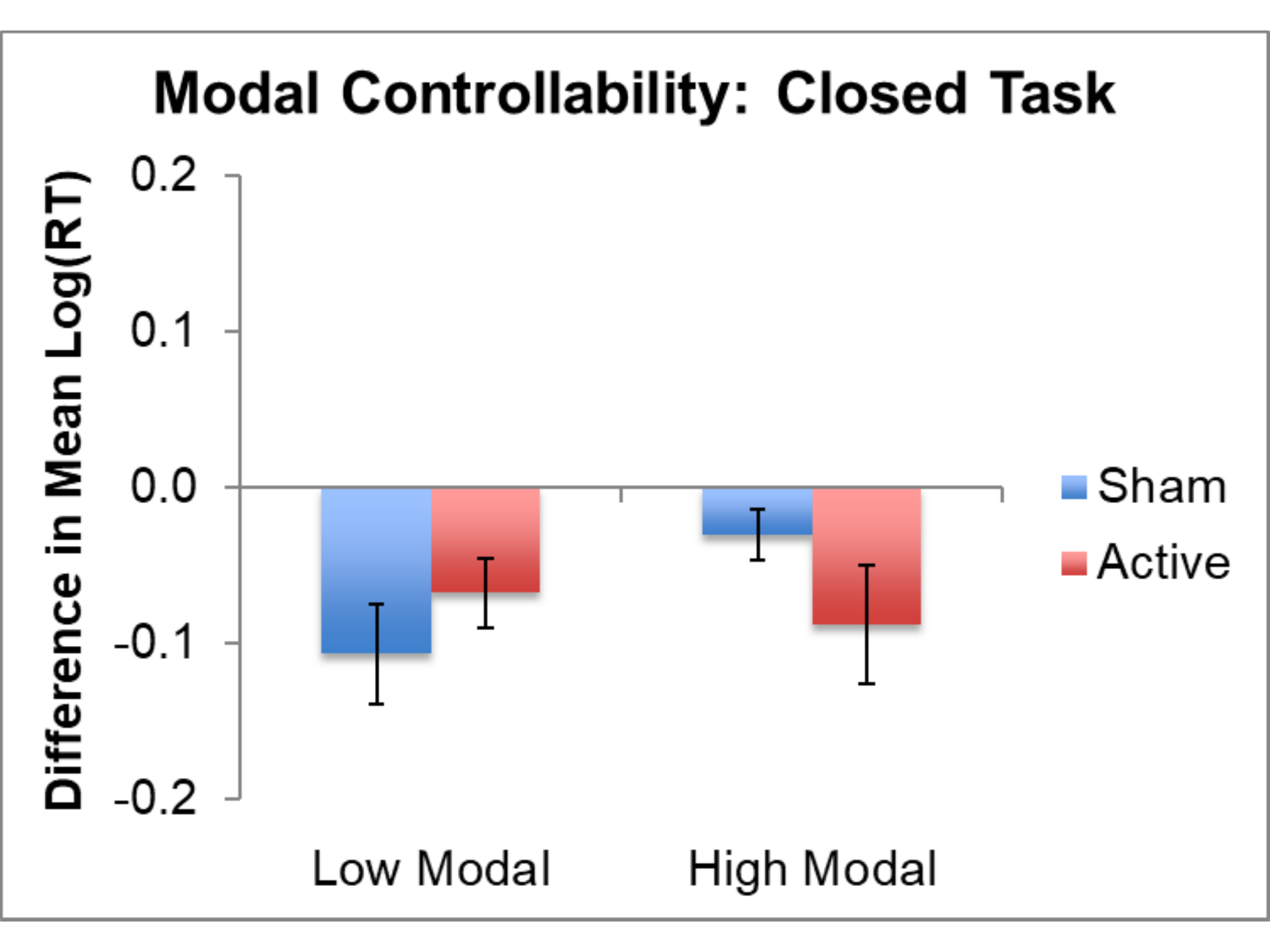}}
		\caption{\textbf{TMS effects in the significant interaction between TMS and modal controllability during the closed language task.} ``High" and ``Low" represent modal controllability ranked values above and below the median for the sample for illustration. Error bars represent standard error (SE) of the differences in mean log-transformed RTs. Lower values on the y-axis represent faster responses in the second session relative to the first session. See Table~\ref{table:modalclosed} for full model results. Note that for transparency, the mean and SE log-transformed RTs do not represent model-adjusted values.}\label{fig:modalclosed}
	\end{figure}

	Modal controllability was also associated with changes in open-ended task performance as a function of TMS. However, in contrast to the boundary controllability findings, individuals with lower (and not higher) modal controllability exhibited practice-related interference effects in the absence of stimulation (i.e., sham group), with slower RTs after (when compared to before) sham stimulation (i.e., mean (and SE) RTs for modal controllability values lower than the median before vs. after = 7.17 (.10) vs. 7.32 (.11), respectively). Here too the practice-related interference effect associated with lower modal controllability was eliminated following cTBS of the LIFG (i.e., similar RTs before and after cTBS; mean (and SE) RTs for modal control values lower than the median before vs. after = 7.13 (.05) and 7.13 (.07), respectively). For individuals with higher modal controllability, performance before vs. after stimulation did not differ as a function of active vs. sham stimulation. One possible explanation for these findings is that reaching difficult states becomes harder with practice -- a processing cost that would likely affect individuals whose LIFG is not particularly well-suited to theoretically drive the network into difficult-to-reach states. Because modal controllability is negatively related with the density of connections to a node, individuals with lower LIFG modal controllability could have difficulty selecting specific, isolated representations that facilitate responses according to varying task demands, such as retrieving and selecting a word appropriate for a given cue. Local inhibition of the LIFG could remove the interference effect in these individuals (see Table~\ref{table:modalopen} and Fig.~\ref{fig:modalclosed}).

	\begin{table}[ht]
		\caption*{\textbf{Mixed effects model for the effects of modal controllability, session, and TMS in the open tasks.}} 
		\begin{tabular}{l c c c c c c} \toprule
			
			{} & {Estimate} & {Std. Error} & {t-value} & {\emph{p}} \\ \midrule
			Intercept  & 7.033 & 0.454 & 15.504 & $<$0.001***\\
			Session    & 0.258& 0.169 & 1.53 & 0.126\\
			ActiveSham & -0.856& 0.885 & -0.967 & 0.333\\	  
			Modal   & 0.002 & 0.007 & 0.315 & 0.753\\	 
			Session*Modal   & -0.004 & 0.003 & -1.523 & 0.128\\	 
			Session*ActiveSham   & 0.977 & 0.329 & 2.971 & 0.003**\\
			Modal*ActiveSham   & 0.013 & 0.014 & 0.954 & 0.340\\	 
			Modal*ActiveSham*Session   & -0.014 & 0.005 & -2.785 & 0.005**\\		 
			\bottomrule	
		\end{tabular}
		\par
		\caption{ActiveSham = active transcranial magnetic stimulation effect relative to sham. Modal = continuous modal controllability effect. Session = effect of the second relative to the first session. *Denotes significance at $p < 0.05$, **denotes significance at $p < 0.01$, and ***denotes significance at $p < 0.001$.}
		\label{table:modalopen}
	\end{table}
	
	\newpage
	
	\begin{figure} 
		\centerline{\includegraphics[width=4in]{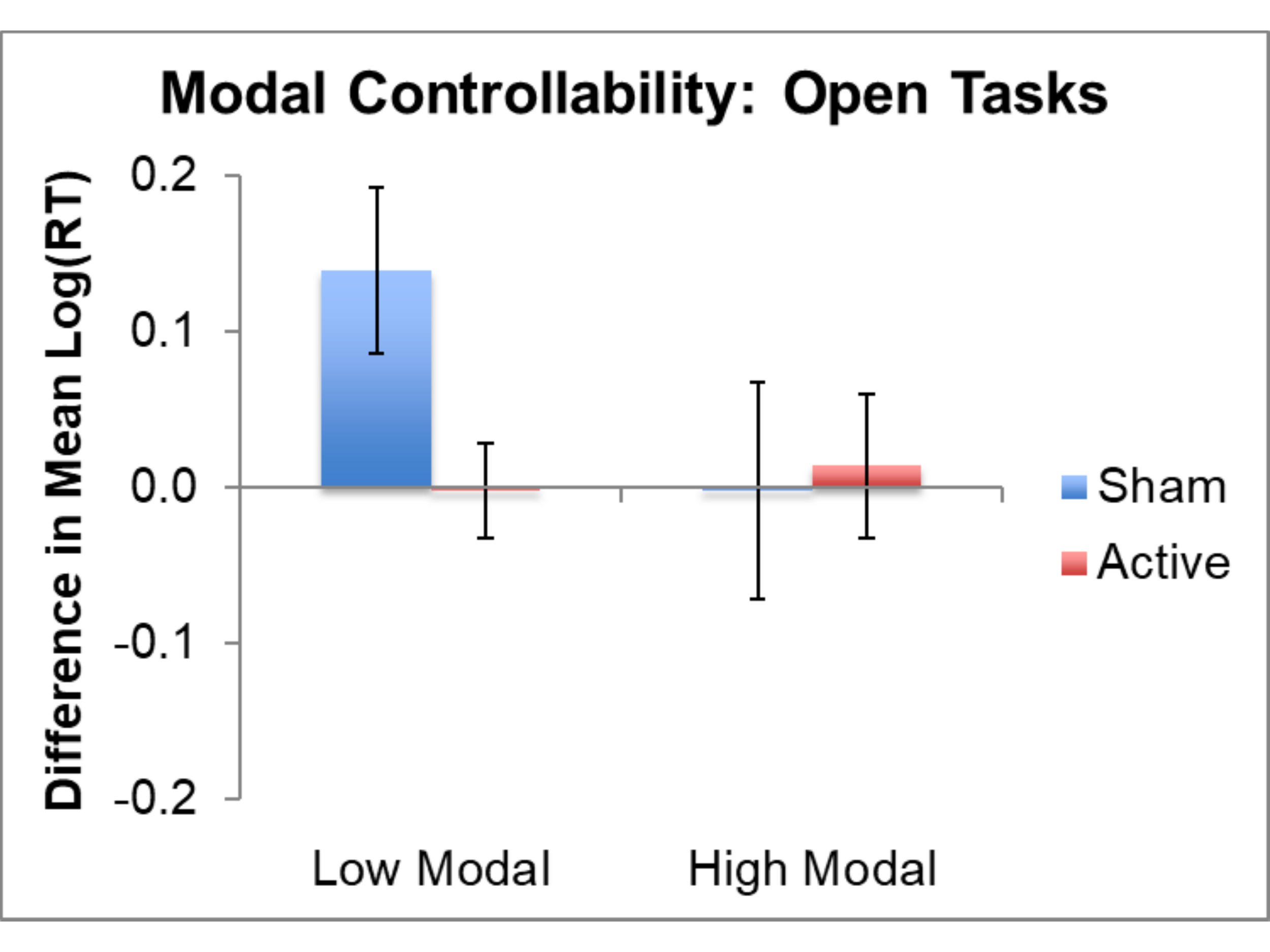}}
		\caption{\textbf{TMS effects in the significant interaction between TMS and modal controllability in open language tasks.} ``High" and ``Low" represent modal controllability ranked values above and below the median for the sample for illustration. Error bars represent standard error (SE) of the differences in mean log-transformed RTs. Lower values on the y-axis represent faster responses in the second session relative to the first session. See Table~\ref{table:modalopen} for full model results. Note that for transparency, the mean and SE log-transformed RTs do not represent model-adjusted values.}\label{fig:modalopen}
	\end{figure}
	\clearpage
	\newpage
	
	\paragraph{\emph{Closed-ended language task performance, TMS, and boundary and modal controllability.}}
	The model assessing boundary controllability effects on performance in the closed-ended task revealed that although participants became faster with practice (i.e., practice-related facilitation) -- an effect that was larger in individuals with low compared to high boundary controllability (i.e., low boundary controllability mean (and SE) RTs before vs. after = 7.11 (.07) vs. 7.00 (.06), respectively; high boundary controllability mean (and SE) RTs before vs. after = 7.00 (.04) vs. 6.95 (.05), respectively) -- this effect was not modulated by cTBS to the LIFG. Thus, the lack of a TMS effect as it relates to boundary controllability and performance on the closed-ended task suggests that the LIFG's ability to integrate and segregate communication between brain networks may not play an important role when the task demands require retrieving and selecting a single, correct representation (rather than one from among several possible acceptable alternatives). (See Table~\ref{table:boundclosed} and Fig.~\ref{fig:boundopen}).
	
	\begin{table}[ht]
		\caption*{\textbf{Mixed effects model for the effects of boundary controllability, session, and TMS in the closed task.}} 
		\begin{tabular}{l c c c c c c} \toprule
			
			{} & {Estimate} & {Std. Error} & {t-value} & {\emph{p}} \\ \midrule
			Intercept  & 7.167 & 0.049 & 145.975 & $<$0.001***\\
			Session    & -0.042& 0.018 & -2.267 & 0.023\\
			ActiveSham & 0.037& 0.098 & 0.383 & 0.702\\	  
			Boundary   & 0.111 & 0.089 & 1.242  & 0.214\\	 
			Session*Boundary   & -0.064 & 0.026 & -2.486  & 0.013\\	 
			Session*ActiveSham   & -0.016 & 0.028 & -0.589  & 0.556\\
			Boundary*ActiveSham   & 0.017 & 0.179 & 0.093  & 0.926\\	 
			Boundary*ActiveSham*Session   & 0.041 & 0.051 & 0.819 & 0.419\\		 
			\bottomrule	
		\end{tabular}
		\par
		\caption{ActiveSham = active transcranial magnetic stimulation effect relative to sham. Boundary = binarized boundary controllability effect. Session = effect of the second relative to the first session. *Denotes significance at $p < 0.05$, **denotes significance at $p < 0.01$, and ***denotes significance at $p < 0.001$.}
		\label{table:boundclosed}
	\end{table}
	
	\newpage
	
	\begin{figure} 
		\centerline{\includegraphics[width=4in]{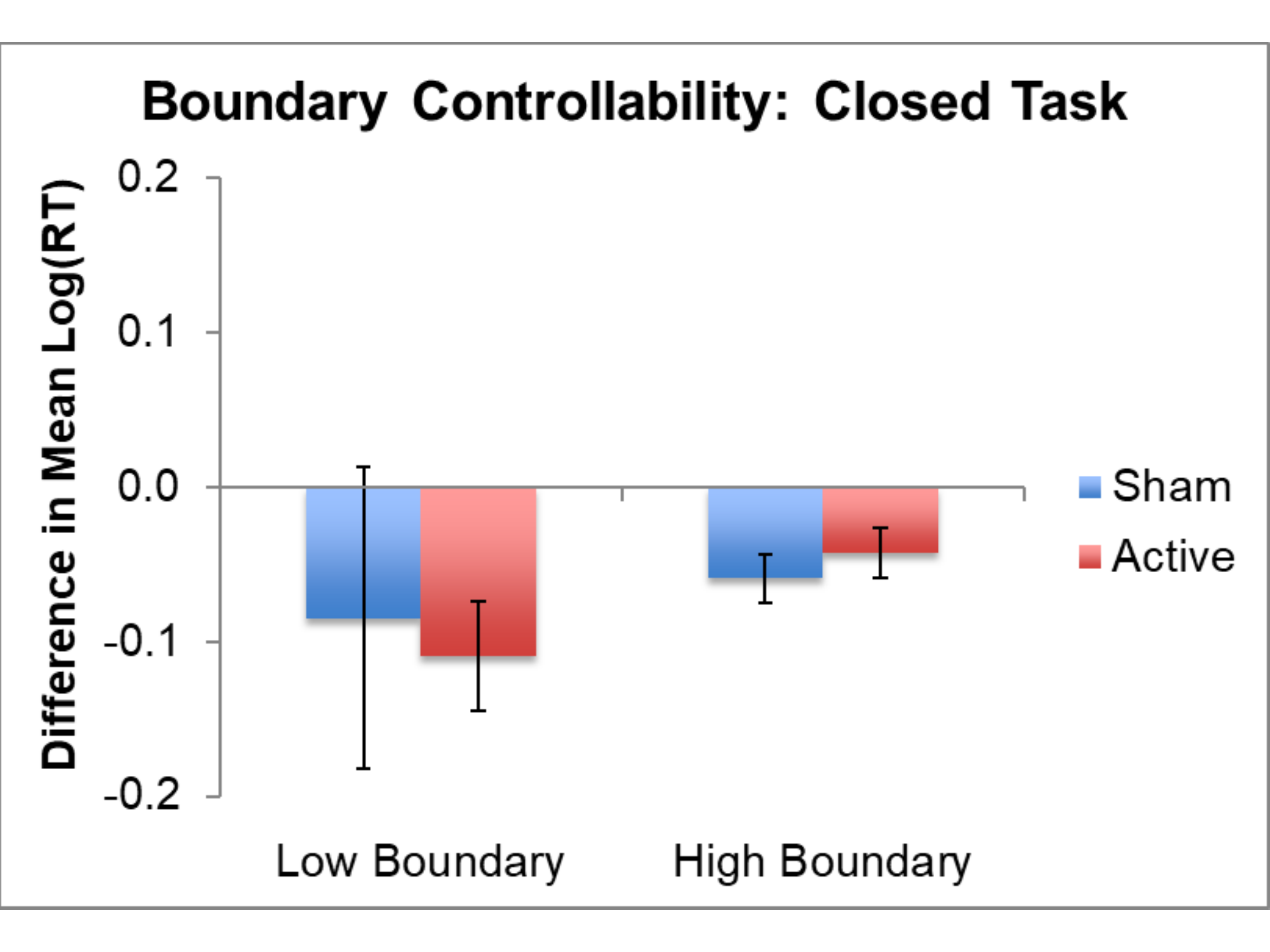}}
		\caption{\textbf{TMS effects in the significant interaction between TMS and boundary controllability during the closed language task.} Error bars represent standard error (SE) of the differences in mean log-transformed RTs. Lower values on the y-axis represent faster responses in the second session relative to the first session. See Table~\ref{table:boundclosed} for full model results. Note that for transparency, the mean and SE of the log-transformed RTs do not represent model-adjusted values.}\label{fig:boundclosed}
	\end{figure}
	\clearpage
	\newpage
	
	By contrast, modal controllability within the LIFG was associated with TMS-induced changes in closed-ended task performance. Relative to sham, practice-related facilitation (mean (and SE) RTs before vs. after = 7.01 (.09) vs. 6.98 (.08), respectively) increased with higher modal controllability (mean (and SE) RTs before vs. after = 7.14 (.08) vs. 7.06 (.07), respectively). Specifically, individuals in which the LIFG serves as a strong modal controller exhibit blunted practice effects that can be enhanced via cTBS. See Table~\ref{table:modalclosed} and Fig.~\ref{fig:modalclosed}.
	
	\begin{table}[ht]
		\caption*{\textbf{Mixed effects model for the effects of modal controllability, session, and TMS during the the closed task.}} 
		\begin{tabular}{l c c c c c c} \toprule
			
			{} & {Estimate} & {Std. Error} & {t-value} & {\emph{p}} \\ \midrule
			Intercept  & 6.243 & 0.530 & 11.774 & $<$0.001***\\
			Session    & 0.302 & 0.160 & 1.890 & 0.059\\
			ActiveSham & 1.706 & 1.033 & 1.651 & 0.099\\	  
			Modal   & 0.013 & 0.008 & 1.518 & 0.129\\	 
			Session*Modal   & -0.006 & 0.003 & -2.361 & 0.018*\\	 
			Session*ActiveSham   & -0.921 & 0.299 & -3.076 & 0.002**\\
			Modal*ActiveSham   & -0.026 & 0.016 & -1.652 & 0.099\\	 
			Modal*ActiveSham*Session   & 0.015 & 0.005 & 3.127 & 0.002**\\		 
			\bottomrule	
		\end{tabular}
		\par
		\caption{ActiveSham = active transcranial magnetic stimulation effect relative to sham. Modal = modal controllability effect. Session = effect of the second relative to the first session. *Denotes significance at $p < 0.05$, **denotes significance at $p < 0.01$, and ***denotes significance at $p < 0.001$.}
		\label{table:modalclosed}
	\end{table}
	
	\newpage
	
	\clearpage
	\newpage
	
	\paragraph*{\emph{Summary of LIFG controllability results.}} To summarize, boundary controllability, representing the theoretical ability to integrate and segregate network communication may represent a control demand specific to open-ended language tasks, namely the need to retrieve and select a single word in the face of competing, alternative words when many responses may be equally appropriate (given a cue). This is suggested by our findings that boundary controllability within the LIFG predicted individuals' responsiveness to TMS in open-ended (but not the closed-ended) language tasks. By contrast, modal controllability -- the ability to steer the network in difficult-to-reach states -- may represent a control demand common to both open- and closed-ended language tasks, i.e. the need to select specific words according to varying task demands. This is suggested by our findings that modal controllability within the LIFG predicted individuals' responsiveness to TMS in both open- and closed-ended tasks. Taken  together, the analyses reveal dissociable control roles of the LIFG relevant for understanding this region's involvement in open- and closed-ended language tasks by virtue of how individuals respond to exogenous brain stimulation.

	\paragraph{\emph{Control comparison examining TMS effects and boundary and modal control within the left lateral occipital cortex.}}
	As a control analysis for LIFG region-specific effects within the connectome, we replicated our multilevel regression analyses using the boundary and modal controllability ranked values from the left lateral occipital region (see Fig.~\ref{fig:controllabilityplot}). In a functional context, the occipital lobe contributes ``bottom up" processing in the context of the ``top down" control processes thought to be mediated by the LIFG. Furthermore, the left lateral occipital region constitutes a strong statistical control, as variability in the rank network control values in the left lateral occipital lobe and LIFG (\emph{pars triangularis}) region were comparable for boundary controllability (i.e., St.D. = 12.8 and 5.4, respectively) and modal control (i.e., St.D. = 12.8 and 16.1, respectively). These analyses indicate that the influence of TMS on open tasks relative to sham does not interact with occipital cortex boundary controllability (specific effect: $p = 0.22$) or modal controllability (specific effect: $p = 0.25$). The influence of TMS on the closed tasks relative to sham also does not interact with occipital cortex boundary controllability (specific effect: $p = 0.16$) or modal controllability (specific effect: $p = 0.23$). Overall, these observations indicate that controllability at the ``top-down" region that was the site of stimulation is distinctly related to session and TMS effects.

	
	\section*{Discussion}
	In this paper, we examined the hypothesis that network controllability in the LIFG is related to language performance in tasks with open- and closed-ended response demands. We explicitly test this hypothesis by linking variability in the vulnerability of controlled language function to perturbation by TMS to LIFG controllability. In this study, we integrate two separately developing theoretical frameworks from cognitive neuroscience and emerging applications of control theory to human brain networks \cite{gu2015controllability}. In cognitive neuroscience, the LIFG is identified as a site that mediates controlled language function; however, the mechanisms by which the LIFG executes this role in brain networks is unknown. Network control theory is postulated to be a useful framework to understand the organization for human cognitive control and performance variability based on the role of anatomical regions in the structural connectome \cite{gu2015controllability,tang2017developmental}. 
	
	To test this experimentally, we constructed structural brain networks from diffusion spectrum imaging data acquired in 28 healthy adult individuals and administered inhibitory TMS (n=19) or sham stimulation (n=9) between two repetitions of language tasks that differ in open- and closed- response demands. We anticipated that dissociable network control roles in the \emph{pars triangularis} of the LIFG -- boundary and modal controllability -- would be associated with distinct changes in the tasks with open and closed demands, respectively. Our results suggest that such a dissociation exists. Specifically, during the open tasks, active TMS in individuals with higher boundary controllability was associated with a reversal of an interference effect, whereas lower modal controllability was associated with a reversal of a learning effect. In contrast, during the closed task, active TMS in individuals with higher modal controllability was associated with an enhanced learning effect.
	
	Importantly, by evaluating LIFG network control roles in the context of performance changes following active as compared to sham stimulation, the current study provides insight into the mechanisms by which the LIFG supports controlled language function. In particular, our finding that boundary controllability within the LIFG was associated with TMS effects in the open-ended, but not closed-ended, language tasks suggests that the integration and segregation of intermodular communication may be particularly important for retrieving words when several alternative words compete for selection, consistent with accounts of IFG mediating retrieval and selection of distributed representations in semantic memory \cite{bedny2008semantic,pulvermuller2013neurons}. Speculatively, if the LIFG is relatively better positioned to control high-level inter-modular communication, increased processing demands may confer more interference with sustained task performance. Thus, inhibition of the LIFG may release accumulating competition among valid responses that results in practice-related interference for open-ended tasks. This speculation is tenable given that the theoretical role of stronger boundary controllers is to facilitate interactions among major brain networks, which could simultaneously confer advantages to general integrated processing demands but disadvantages as competition between semantically activated representations accumulates.
	
	By contrast, modal controllability within the LIFG was associated with TMS effects in both open- and closed-ended language tasks, indicating that modal controllability mediates controlled aspects of language production independent  of the open/closed response demand distinction. This observation suggests that the LIFG's theoretical ability to drive the network into difficult-to-reach states \cite{pasqualetti2014controllability,gu2015controllability} may partially mediate the retrieval and selection of task appropriate responses according to varying demands, consistent with some network-level accounts for these processes \cite{friederici2015grounding,ye2014brain,fedorenko2014reworking}. Thus, for open tasks in which competition may accumulate with practice, increasing practice-related interference with lower modal controllability may reflect increased difficulty in achieving rare states as interference accumulates. This effect is ameliorated with inhibition of the LIFG, potentially because TMS may partially suppress the LIFG's role in higher processing demands \cite{nicolo2016neurobiological,vuksanovic2015improvement} due to an increased number of connections within the network (i.e., lower modal controllability associated with denser connections). During the closed task, blunted practice effects in individuals with higher modal controllability can be improved with cTBS. Speculatively, this could be because higher modal controllability induces interference due to serially retrieving specific responses (e.g., via semantic satiation) that is ameliorated via TMS. However, there are potential alternative explanations for this finding due to the complex retrieval and selection demands that may influence the accumulation of interference across sessions within the tasks. Nevertheless, cTBS modulated the relationship between varying modal controllability within the LIFG and performance on open- and closed-ended response demands. This indicates that this control role mediates controlled aspects of language production regardless of whether there are one or many correct responses.
	
	Collectively, these findings provide a basis for integrating the local influences of TMS, mediating anatomical organization across the brain, and performance on controlled language tasks. Speculatively, our results could promote work that aims to connect neurophysiological and network neuroscience \cite{medaglia2015cognitive,bassett2017network} more generally. While the LIFG may be related to control functions generally \cite{brass2005role}, it may be possible to pair TMS, diffusion imaging, and task manipulations to dissociate specific contributions outside the language domains. In addition, the executive processes involved in language may not be unique to the frontal lobe \cite{whitney2010neural}, and whether or not similar network controllers in other parts of the brain influence controlled language function remains to be seen.
	
	It is important to note that presumably GABA-mediated inhibition \cite{trippe2009theta} is a consequence of the direct effects of rTMS and a potential mediating mechanism for variability in word selection processes \cite{snyder2010neural}. However, this local mechanism is situated in the context of individually variable anatomical networks, which our current work establishes are relevant to dissociable control processes in the brain associated with variability in language task performance. Some cross-sectional work illustrates the link between anatomical network controllability and fMRI dynamics across neurodevelopment \cite{tang2017developmental}. While the current study used a simplified model of dynamics that has been demonstrated to predict the controllability of Wilson-Cowan \cite{muldoon2016stimulation} and Kuramoto \cite{tiberi2017synchronization} oscillators coupled by empirically measured anatomical brain networks, it is important to note that theoretical predictions about controllability would be further strengthened by evaluating empirically measured neural activity in response to exogenous brain stimulation. For example, demonstrating that integrated or segregated or difficult to reach BOLD or EEG states are influenced by TMS as a function of boundary and modal controllability, respectively, would support the theoretical notions described here. Validating dynamic predictions with concurrent or post-TMS data will be crucial to understand the dynamic shifts responsible for behavior change.
	
	Future studies could examine larger cohorts, including the effects of TMS at the LIFG over broader age ranges and in patients with neuropsychiatric conditions. In addition, while the current results establish a link between TMS boundary controllability and response times during controlled language function, they are not specific to classically examined selection or retrieval demands at the item level. We could thus examine the interactions between the LIFG and specific controllable subnetworks of the brain involved in more general or specific control processes \cite{fedorenko2014reworking}, and different behavioral task designs such as open-ended number generation and closed-ended sentence paradigms to examine relationships between network controllability and item selection and retrieval demands. Finally, we applied a theta-burst stimulation sequence, but numerous other stimulation procedures have been used to influence cognitive-emotional functioning, including in the LIFG. Future studies could use varying stimulation parameters to examine sensitivity of controlled language function to different stimulation intensities, as well as their interaction with network controllability. Finally, while we used methods consistent with earlier work applying controllability analysis to anatomical networks \cite{gu2015controllability,betzel2016optimally,tang2017developmental,gu2017optimal}, future studies could investigate the optimal parameters for predicting TMS effects including selection of fiber tracking, parcellation, and network weighting or binarizing procedures.
	
	By examining the relationship between inter-individual variability in LIFG  controllability and controlled language function before and after brain stimulation, we establish a bridge between a neuroscientist's notion of a controlled language process and an engineer's notion of network control. The current results demonstrate that linking network controllability in white matter networks with experimental manipulation involving TMS can reveal associations between regional network control roles and cognitive susceptibility to brain stimulation. It is possible to identify dissociable control roles of a single region of the brain using simplified dynamic models in anatomical connectivity. Similar experiments may elucidate the role of the LIFG in specific and general cognitive control functions in the human connectome.

	\newpage
	\section*{Acknowledgments}
	JDM acknowledges support from the Office of the Director at the National Institutes of Health through grant number 1-DP5-OD-021352-01. DSB acknowledges support from the John D. and Catherine T. MacArthur Foundation, the Alfred P. Sloan Foundation, the Army Research Laboratory and the Army Research Office through contract numbers W911NF-10-2-0022 and W911NF-14-1-0679, the National Institute of Mental Health (2-R01-DC-009209-11), the National Institute of Child 415 Health and Human Development (1R01HD086888-01), the Office of Naval Research, and the National Science Foundation (BCS-1441502, BCS-1430087, and CAREER PHY-1554488). JDM, DSB, and RHH acknowledge support from a Translational Neuroscience Initiative through the Perelman School of Medicine at the University of Pennsylvania.

	\clearpage
	\newpage
	
	\bibliographystyle{jneurosci}
	\bibliography{JDMReferences}

\end{document}